\documentclass{article}
\usepackage{hq2kpro}
\usepackage{epsfig}
\usepackage{colordvi}

\newcommand{\pt}{$p^{}_T$}

\newcommand{\dpimumu}{$D^+\rightarrow\pi^+\mu^+\mu^-$}

\newcommand{\dkee}{$D^+\rightarrow K^+ e^+ e^-$}
\newcommand{\dkmunu}{$D^+\rightarrow\overline{K}^{\,0}\mu^+\nu$}
\newcommand{\dkpimue}{$D^0\rightarrow K^-\pi^+\mu^- e^+$}
\newcommand{\dmumu}{$D^0\rightarrow\mu^+\mu^-$}
\newcommand{\dee}{$D^0\rightarrow e^+e^-$}
\newcommand{\dskmumu}{$D^+_s\rightarrow K^+\mu^+\mu^-$}
\newcommand{\dphimumu}{$D^0\rightarrow\phi\,\mu^+\mu^-$}

\newcommand{\dkpi}{$D^0\rightarrow K^-\pi^+$}
\newcommand{\dpipi}{$D^0\rightarrow\pi^+\pi^-$}
\newcommand{\dkpp}{$D^+\rightarrow K^-\pi^+\pi^+$}
\newcommand{\dppp}{$D^+\rightarrow\pi^-\pi^+\pi^+$}
\newcommand{\dsphipi}{$D^+_s\rightarrow \phi\,\pi^+$}

\newcommand{\dkstarpp}{$D^0\rightarrow\overline{K}^{*0}\pi^+\pi^-$}
\newcommand{\dphipp}{$D^0\rightarrow \phi\,\pi^+\pi^-$}
\newcommand{\drhopp}{$D^0\rightarrow\rho^0\pi^+\pi^-$}
\newcommand{\dkppp}{$D^0\rightarrow K^-\pi^+\pi^-\pi^+$}
\newcommand{\dpppp}{$D^0\rightarrow\pi^+\pi^-\pi^+\pi^-$}

\newcommand{\dll}{$D^0\rightarrow\ell^+_1\ell^-_2$}
\newcommand{\dpll}{$D^+\rightarrow P\,\ell^{}_1\ell^{}_2$}

\newcommand{\dspll}{$D^+_s\rightarrow P\,\ell^{}_1\ell^{}_2$}
\newcommand{\dorspll}{$D^+_{(s)}\rightarrow P\,\ell^{}_1\ell^{}_2$}
\newcommand{\dvll}{$D^0\rightarrow V\ell^+_1\ell^-_2$}
\newcommand{\dkstarll}{$D^0\rightarrow\overline{K}^{*0}\ell^+_1\ell^-_2$}
\newcommand{\dphill}{$D^0\rightarrow\phi\,\ell^+_1\ell^-_2$}
\newcommand{\drholl}{$D^0\rightarrow\rho^0\ell^+_1\ell^-_2$}
\newcommand{\dhhll}{$D^0\rightarrow h^{}_1 h^{}_2\ell^{}_1\ell^{}_2$}
\newcommand{\dppll}{$D^0\rightarrow\pi\pi\ell^{}_1\ell^{}_2$}
\newcommand{\dkpll}{$D^0\rightarrow K\pi\ell^{}_1\ell^{}_2$}
\newcommand{\dkkll}{$D^0\rightarrow KK\ell^{}_1\ell^{}_2$}
\newcommand{\dkkpp}{$D^0\rightarrow K^+K^-\pi^+\pi^-$}

\newcommand{\geve}{~GeV}
\newcommand{\gevp}{~GeV/$c$}
\newcommand{\gevm}{~GeV/$c^2$}

\newcommand{\mevm}{~MeV/$c^2$}

\newcommand{\AmS}{{\protect\the\textfont2
  A\kern-.1667em\lower.5ex\hbox{M}\kern-.125emS}}

\begin{document}
\title{ 
\vspace*{-1.0in}
\begin{flushright}
{\normalsize UCTP-113-00}
\end{flushright}
\vspace*{0.30in}
SEARCHES FOR RARE AND FORBIDDEN DECAYS OF CHARM:
RECENT RESULTS FROM FNAL
}
\author{
A.\ J.\ Schwartz \\
{\em  University of Cincinnati, Cincinnati, Ohio 45221 USA}
}
\maketitle
\baselineskip=11.6pt
\begin{abstract}
We review results on rare and forbidden decays of $D^0$, $D^+$,
and $D^+_s$ mesons from experiments at FNAL. The decay modes
studied have two leptons in the final state and, if observed,
would constitute evidence for flavor-changing neutral-current,
lepton-flavor-violating, or lepton-number-violating processes. 
To date, no evidence for these decays has been observed and upper 
limits are obtained for their branching fractions. These limits can 
constrain various extensions to the Standard Model. We present new 
upper limits from FNAL E791 on the branching fractions for more 
than two dozen three- and four-body decay modes.
\end{abstract}
\baselineskip=14pt

\section{Introduction}

\begin{figure}[thb]
\begin{center}
\hbox{
\epsfig{file=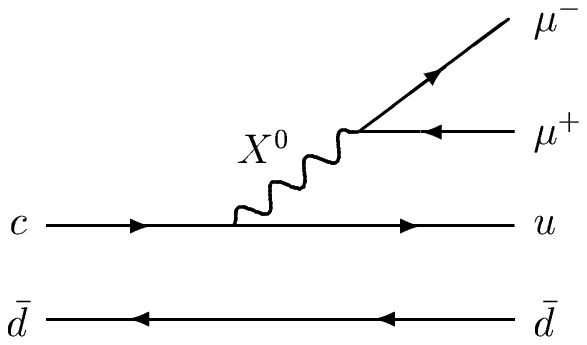,height=1.25in}
\hspace*{0.20in}
\epsfig{file=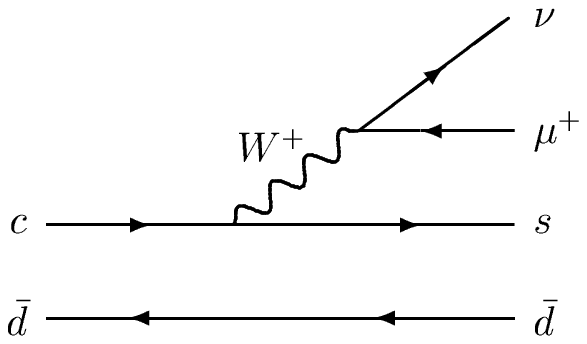,height=1.25in}
}
\end{center}
\vspace*{-0.40in}
\caption{\it Nonstandard flavor-changing neutral-current
decay (left) and Standard Model charged-current decay (right).
\label{fig:feynman}}
\end{figure}

Searches for rare and forbidden decays of charm are concerned 
with final states containing two charged leptons. Such processes 
occur via flavor-changing neutral-current amplitudes, 
lepton-flavor-violating amplitudes (leptons belonging to 
different families), or lepton-number-violating amplitudes 
(leptons belonging to one family but having the same sign charge). 
Diagrams for these amplitudes typically contain new types of 
particles having high masses; thus, these decays probe energy 
scales which cannot be accessed directly. For example, the amplitude 
for the flavor-changing neutral-current decay \dpimumu\ is expected 
to be proportional to $\tilde{g}^2/M^2_X \times ({\rm phase\ space})$, 
where $\tilde{g}$ is a coupling constant and $M^{}_X$ is the mass
of some unknown propagator (see Fig.~\ref{fig:feynman}). 
The amplitude for the Standard Model decay \dkmunu\ is 
proportional to $g^2/M^2_W \times ({\rm phase\ space})$, and thus 
$\Gamma(\pi^+\mu^+\mu^-)/\Gamma(\overline{K}^{\,0}\mu^+\nu) =
\tilde{g}^4M^4_W/(g^4M^4_X)\times ({\rm phase\ space\ ratio})$.
If $\tilde{g}\approx g$, then
\begin{equation}
M^{}_X \ \approx\  M^{}_W\left[\frac{B(D^+\rightarrow\overline{K}^{\,0}\mu^+\nu)} 
{B(D^+\rightarrow\pi^+\mu^+\mu^-)}\times ({\rm phase\ space\ ratio})\right]^{1/4}\,.
\end{equation}
Inserting numbers one finds that a branching fraction 
$B(D^0\rightarrow\pi^+\mu^+\mu^-)\approx$\,$10^{-5}$ 
corresponds to a mass $M^{}_X\approx 700$\gevm.

Over the past year, FNAL E791 has published new limits on branching
fractions for 24 different two- and three-body decay modes\cite{e791:3body}, 
and submitted for publication new limits for 27 additional three- and 
four-body decay modes\cite{e791:4body}. In most cases
the E791 limits are the most stringent available. Here we 
review these results and also briefly discuss competitive 
results from other FNAL experiments.

\section{The E791 Experiment}

FNAL E791\footnote{The collaboration consists of:
CBPF (Brazil), Tel Aviv, CINVESTAV (Mexico), Puebla (Mexico),
U.~C.\ Santa Cruz, Cincinnati, Fermilab, Illinois Institute of 
Technology, Kansas State, Massachusetts, Mississippi, Princeton, 
South Carolina, Stanford, Tufts, Wisconsin, and Yale.} is a 
hadroproduction experiment studying the weak decays of charm mesons 
and baryons. The charm particles were produced by impinging a 
500~GeV/$c$ $\pi^-$ beam on five thin target foils.  The most upstream 
foil consisted of platinum; the other foils consisted of carbon
(diamond). All foils were separated by about 15~mm such that $D$ mesons 
decayed predominately in the air gaps between foils. The experimental 
apparatus\cite{e791:epj} consisted of a silicon vertex detector followed 
by a two-magnet spectrometer, two segmented Cerenkov counters for hadron 
identification, an electromagnetic calorimeter for electron identification, 
and iron shielding followed by scintillator counters for muon identification.
The downstream silicon vertex detector consisted of 17 planes of silicon 
and was used to reconstruct decay vertices downstream of the interaction 
vertex. The spectrometer consisted of 35 planes of drift chambers
and two proportional wire chambers. The two dipole magnets bent 
particles in the horizontal plane and had \pt\ kicks of +210~GeV/$c$ 
and +320~GeV/$c$. The Cerenkov counters contained gases with different
indices of refraction; together they provided $\pi/K/p$ discrimination
over the momentum range 6--60~GeV/$c$. 
Data were recorded using a loose transverse energy trigger. After 
reconstruction, events with evidence of well-separated interaction 
and decay vertices were retained for further analysis. The experiment 
took data from September, 1991 to January 1992, recording the world's 
largest sample of charm decays at that time. The final number of
reconstructed decays is over 200\,000.

\section{Event Selection}

E791 has searched for two-, three-, and four-body $D^0$ decays 
such as \dee, \dphimumu, and \dkpimue; and three-body $D^+$ and 
$D^+_s$ decays such as \dpimumu\ and \dskmumu. Here and throughout 
this paper, charge-conjugate modes are included unless otherwise noted. 
The sensitivity of a search is determined (or normalized) by counting 
events in a topologically-similar hadronic decay channel such as \dkpp\ 
for which the branching fraction is known. For all searches, the event 
selection proceeded via a ``blind analysis'' technique in order to avoid 
biasing the choice of selection criteria. This technique has three steps:
(a)~all events having a reconstructed mass within a mass window or ``box''
around $m^{}_D$ are removed from the sample;
(b)~the selection criteria are chosen by optimizing the ratio $S/\sqrt{B}$, 
where $S$ is the number of signal events from a Monte Carlo simulation 
that pass all criteria, and $B$ is the number of events from data that 
are within a ``background box'' which is near --\,but exclusive of\,-- 
the signal box;
(c)~the finalized selection criteria are applied to the events
within the signal box to see if any candidate events remain.
The selection criteria resulting from this procedure are listed in 
Tables~\ref{tab:qualitycuts} and \ref{tab:pidcuts}. The most important 
criterion is that of $SDZ$, which is defined as the distance between the 
interaction vertex and the decay vertex divided by the error in this 
quantity. Values used for this criterion were 12 for $D^0$ and $D^+_s$ 
decays and 20 for longer-lived $D^+$ decays.

\begin{table}[t]
\centering
\caption{ \it E791 selection criteria based on tracking and vertexing.}
\vskip 0.1 in
\renewcommand{\arraystretch}{1.2}
\begin{tabular}{|l|c|}
\hline
{\bf Selection criteria} & {\bf Value} \\
\hline
$SDZ\ \equiv$ & \\
\hspace*{0.20in} $(z^{}_{\rm dec}-z^{}_{\rm int})/
\sqrt{\sigma^2_{\rm dec} + \sigma^2_{\rm int}}$ & 
$>$\,20/12/12  \hspace*{0.10in} ($D^+/D^+_s/D^0$) \\
$min|z^{}_{\rm dec} - z^{}_{\rm target\ edge}|/\sigma^{}_{\rm sec}$ & $> 5$ \\
$z^{}_{\rm dec}- z^{}_{\rm last\ target}$ & $<$\,16~mm \\
$\chi^2_{\rm track}$ & $< 5$ \\
$\chi^2_{\rm vertex}$ & $< 6$ \\
$D$ impact parameter (i.p.) &   \\
\hspace*{0.40in} w/r/t int. vertex & 
$< 30/40$~$\mu$m \hspace*{0.05in} ($D^0$ 3-body,\,4-body/others) \\
$\prod_{\rm tracks}\left(\frac{\rm i.p.\ w/r/t\ decay\ vertex}
{\rm i.p.\ w/r/t\ int.\ vertex}\right)$ & 
$<$\,0.01/0.001/0.0005 \hspace*{0.05in}  (2-body/3-body/4-body) \\
\pt\ (transverse to $D$ direction) & $< 0.20/0.25/0.30$~GeV/$c$ 
\hspace*{0.10in}  ($D^+/D^+_s/D^0$) \\
$t\ \equiv\ m^{}_D\,\times\,(z^{}_{\rm dec}-z^{}_{\rm int})/p$ & 
$<$\,5/3/(3 or 2.5)~ps  \hspace*{0.10in}  ($D^+/D^+_s/D^0$) \\
\hline
\end{tabular}
\label{tab:qualitycuts}
\end{table}

\begin{table}[t]
\centering
\caption{ \it E791 selection criteria based on particle identification.}
\vskip 0.1 in
\renewcommand{\arraystretch}{1.2}
\tabcolsep=4.5pt
\begin{tabular}{|ll|}
\hline
{\bf Electron:} & EMPROB\,$>90$ (calorimeter response
consistent with $e$) \\
\hline
{\bf Muon:} & 	 
\begin{tabular}{l}
(1)\ $p^{}_\mu> 8$~GeV/$c$ \\
(2)\ $y$-paddle within 1\,$\sigma$ of track projection is hit \\
(3)\ $x$-position as calculated from $y$-paddle TDC \\
\hspace*{0.35in} time is within 60~cm of track projection \\
(4)\ If $y$-paddle within 1\,$\sigma$ of track projection is {\it not\/} hit, \\
\hspace*{0.35in} then $x$-paddle to which track projects is hit \\
\end{tabular}  \\
\hline
{\bf Kaon:} & 	 
$\begin{array}{ccl}
P^{}_{\rm \check{C}erenkov}
 & > & 0.10\ \ (D^+\rightarrow K\pi\pi {\rm\ norm.\ for\ } 
				D^+_{(s)}\rightarrow\pi\ell\ell\,) \\
 & > & 0.13\ \ (D^0\rightarrow K\pi {\rm\ norm.\ for\ } 
				D^0\rightarrow\ell^+\ell^-, \\
 &  & \hspace*{0.40in} D^0\rightarrow K\pi\pi\pi {\rm\ norm.\ for\ } 
				D^0\rightarrow\pi\pi\ell\ell, \\
 &  & \hspace*{0.40in} D^+\rightarrow K\ell\ell, \\
 &  & \hspace*{0.40in} D^+\rightarrow K\pi\pi {\rm\ norm.\ for\ } 
				D^+\rightarrow K\ell\ell\,) \\
 & > & 0.18\ \ (D^+_s\rightarrow K\ell\ell, \\
 &  & \hspace*{0.40in} D^+_s\rightarrow\phi\,\pi {\rm\ norm.\ for\ } 
				D^+_s\rightarrow K\ell\ell, \\
 &  & \hspace*{0.40in} D^0\rightarrow K\pi\ell\ell, \\
 &  & \hspace*{0.40in} D^0\rightarrow K\pi\pi\pi {\rm\ norm.\ for\ } 
				D^0\rightarrow K\pi\ell\ell\,) 
\end{array}$   \\
\hline
\end{tabular}
\label{tab:pidcuts}
\end{table}

The mass windows chosen for the various searches depend upon the
number of particles in the final state and on whether any of them are 
electrons. For final states {\it not\/} containing electrons, the 
windows extended $\pm$\,35, $\pm$\,30, or $\pm$\,20\mevm\ around 
$m^{}_D$ for $D^0$, $D^+$, and $D^+_s$ decays, respectively. For 
final states containing one or more 
electrons, the mass windows were asymmetric, extending 40--70\mevm\ 
farther below $m^{}_D$ than above it as the mass distributions 
had low-energy tails resulting from final-state bremsstrahlung.
For search modes containing a $\rho^0$, $\overline{K}^{*0}$, or
$\phi$ meson in the final state, it was required that 
$|m^{}_{\pi^+\pi^-} - m^{}_\rho |< 150$\mevm,  
$|m^{}_{K^-\pi^+} - m^{}_{K^*} |< 55$\mevm, or 
$|m^{}_{K^+K^-} - m^{}_\phi |< 10$\mevm, respectively.

After all selection criteria were applied, no significant excess 
of events above the estimated background was observed. The experiment 
thus sets upper limits as follows:
\begin{equation}
UL(D\rightarrow X) = \left(\frac{N^{}_X}{N^{}_{\rm norm}}\right)
\left(\frac{\varepsilon^{}_{\rm norm}}{\varepsilon^{}_X}\right)\times
B^{}_{\rm norm}\,,
\end{equation}
where $N^{}_X$ is the 90\% C.L. upper limit on the mean number of 
signal events as determined from the number of candidate events 
observed and the estimated background; $N^{}_{\rm norm}$ is the 
number of events observed (after background subtraction) in a 
hadronic normalization channel such as \dkpp;
$\varepsilon^{}_X$ and $\varepsilon^{}_{\rm norm}$ are the overall 
detection efficiencies for the search channel ($D\rightarrow X$) 
and normalization channel, respectively; and $B^{}_{\rm norm}$ 
is the branching fraction for the normalization 
channel as taken from the Particle Data Book\cite{PDG}. 
The upper limits $N^{}_X$ are calculated using the method
of Feldman and Cousins\cite{FeldmanCousins} in order to account 
for estimated background. They are subsequently increased via 
the prescription of Cousins and Highland\cite{CousinsHighland} 
to account for systematic errors. 

There were eight hadronic decay channels used for normalization.
These channels are listed in Table~\ref{tab:normalization} along with 
the number of events obtained for each after background subtraction. 
In general, a dilepton search mode was normalized to a Cabibbo-favored 
hadronic mode having the same number of tracks in the final state and, 
whenever possible, the same daughter particles except for the substitution
of pions for leptons. For example, the $D^+\rightarrow K\ell^{}_1\ell^{}_2$ 
searches were normalized to \dkpp\ decays. However, the \drholl\ searches
were normalized to \dpppp\ decays, as there is no published branching 
fraction for \drhopp.

\begin{table}[t]
\centering
\caption{ \it Hadronic decay modes used to normalize the
E791 searches for \dll, \dorspll, \dvll, and \dhhll\ decays
($P$\,=\,pseudoscalar, $V$\,=\,vector).}
\vskip 0.1 in
\renewcommand{\arraystretch}{1.2}
\begin{tabular}{|llc|}
\hline
{\bf Search mode} & {\bf Normalization mode} & {\bf {\boldmath $N^{}_{\rm norm}$}} \\
\hline
\dll		& \dkpi		& 25210\,$\pm$\,179 \\
\hline
\dpll		& \dkpp		& 24010\,$\pm$\,166 \\
\dspll		& \dsphipi      & 782\,$\pm$\,30 \\
\hline
\drholl		& \dpppp 	& 2049\,$\pm$\,53 \\
\dkstarll	& \dkstarpp	& 5451\,$\pm$\,72 \\
\dphill		& \dphipp	& 113\,$\pm$\,19 \\
\dppll		& \dpppp	& 2049\,$\pm$\,53 \\
\dkpll		& \dkppp	& 11550\,$\pm$\,113 \\
\dkkll		& \dkkpp	& 406\,$\pm$\,41 \\
\hline
\end{tabular}
\label{tab:normalization}
\end{table}

\section{Background Estimate}

There were two main sources of background in E791: ``reflection'' 
background arising from fully-reconstructed hadronic $D$ decays 
in which two of the tracks were misidentified
as leptons, and ``combinatoric'' background arising 
from accidental combinations of tracks and vertices. 
Most of the reflection background was eliminated by
excluding events with invariant masses (assuming all 
daughters to be $\pi$ or $K$) near $m^{}_D$; i.e.,
it was required that 
$|m(h^{}_1h^{}_2h^{}_3h^{}_4) - m^{}_{D^0}| > 35$\mevm,
$|m(h^{}_1h^{}_2h^{}_3) - m^{}_{D^+}| > 30$\mevm, and
$|m(h^{}_1h^{}_2h^{}_3) - m^{}_{D^+_s}| > 20$\mevm.
These requirements were imposed for all
$h^{}_1$\ldots$h^{}_4$ final states listed in 
Table~\ref{tab:background}, with one exception: those
final states having the same number of kaons as that 
of the search mode were not excluded in this manner, as 
the acceptance loss for signal events would have been 
excessive. Instead, background from these modes was 
estimated as follows.

\begin{table}[t]
\centering
\caption{ \it Hadronic decay channels contributing
reflection background to the E791 \dll, \dorspll, \dvll, 
and \dhhll\ samples ($P$\,=\,pseudoscalar, $V$\,=\,vector).}
\vskip 0.1 in
\renewcommand{\arraystretch}{1.2}
\begin{tabular}{|c|c|}
\hline
{\bf Cabibbo-favored} & {\bf Cabibbo-suppressed} \\
\hline
\dkpi\ 	& \dpipi\ 	\\
\hline
$D^+\rightarrow K^-\pi^+\pi^+$ & $D^+\rightarrow \pi^-\pi^+\pi^+$  \\
$D^+_s\rightarrow \pi^-\pi^+\pi^+$ & $D^+\rightarrow K^- K^+\pi^+$  \\
$D^+_s\rightarrow K^- K^+\pi^+$ & $D^+_s\rightarrow K^+\pi^+\pi^-$  \\
$\Lambda^+_c\rightarrow pK^-\pi^+$ &  \\ 
\hline
\dkppp\ & \dpppp\ 	\\
	& \dkkpp\ 	\\
\hline
\end{tabular}
\label{tab:background}
\end{table}

First, the probability for a pair of pions to be misidentified 
as $\mu\mu$, $\mu e$, or $ee$ was estimated from data. This 
probability was then multiplied by the number of events observed 
in the hadronic channel for which the reflection cut could not be 
applied. A factor was included to account for the fraction of these 
events that would reflect into the signal mass window if two 
daughter pions were misidentified as leptons. For example, the 
reflection background in the \dpimumu\ sample arising from 
\dppp\ decays is calculated as 
$P^{}_{\pi\pi\rightarrow\mu\mu}\times 
		N^{}_{\pi^-\pi^+\pi^+}\times f\times J^{}_c$,
where $f$ is the fraction of \dppp\ decays that would reflect into 
the $D^+$ mass window if two daughter pions were misidentified 
as muons (75\% in this example). The factor $J^{}_c$ accounts 
for the different ways a hadronic mode can be misidentified 
as a dilepton mode; e.g., there are two ways that \dppp\ can 
be misidentified as \dpimumu.

The misidentification probabilities for three-body decays were 
obtained from the final $D^+\rightarrow K^-\ell^+_1\ell^+_2$ 
samples, and those for four-body decays from the final 
$D^0\rightarrow K^-\pi^+\ell_1^-\ell_2^+$ samples, where 
all candidates observed (after subtracting combinatoric 
background estimated from mass sidebands) were assumed to 
originate from \dkpp\ and \dkppp\ decays, respectively. 
For example, there were 13, 5.2, and 6 events passing all 
selection criteria for the $K^-\mu^+\mu^+$, $K^-\mu^+ e^+$, 
and $K^- e+ e^+$ samples, respectively, and 17\,730 events 
(after background subtraction) in the \dkpp\ sample. Thus, 
the misidentification probabilities for three-body decays 
were $P^{}_{\mu\mu}=(7.3\,\pm\,2.0)\times$\,$10^{-4}$, 
$P^{}_{\mu e}=(2.9\,\pm\,1.3)\times 10^{-4}$, 
and $P^{}_{ee}=(3.4\,\pm\,1.4)\times 10^{-4}$. 
The reflection backgrounds for the 
$D^+\rightarrow K^-\ell_1^+\ell_2^+$ and
$D^0\rightarrow K^-\pi^+\ell_1^-\ell_2^+$ modes 
themselves were taken to be zero as there was no
independent estimate of the misidentification 
rates. This results in conservative upper limits. 
Unfortunately, there were too many events observed 
for the $D^+\rightarrow K^-\ell_1^+\ell_2^+$ modes
(after accounting for combinatoric background -- see 
below) to set meaningful upper limits.

After the mass reflection cuts, combinatoric background was
estimated by averaging the number of events in the mass sidebands 
both above and below the $D$ signal mass window and scaling this 
number by the size of the signal mass window relative to that of 
the mass sidebands. If there were no events in the higher mass 
sideband, it was assumed that there were no combinatoric background 
events in the signal box. This assumption avoids overestimating 
background due to statistical fluctuations. Because it tends to 
underestimate background, it results in a conservative upper 
limit.

There were also small backgrounds to the
$D^0\rightarrow K^-\pi^+\ell^+\ell^-$, 
$D^0\rightarrow K^+ K^-\ell^+\ell^-$, 
$D^0\rightarrow\overline{K}^{*0}\ell^+\ell^-$, 
and $D^0\rightarrow\phi\,\ell^+\ell^-$ samples
arising from 
$D^0\rightarrow K^-\pi^+\rho^0$, 
$D^0\rightarrow K^+ K^-\rho^0$, 
$D^0\rightarrow\overline{K}^{*0}\rho^0$, and 
$D^0\rightarrow\phi\/\rho^0$ decays,
respectively, where $\rho^0\rightarrow\ell^+\ell^-$.
These backgrounds were estimated using the branching 
fractions for $D^0\rightarrow h^{}_1 h^{}_2\rho^0$,
$D^0\rightarrow V\rho^0$, and $\rho^0\rightarrow\ell^+\ell^-$ 
from the Particle Data Book\cite{PDG}.

\section{Upper Limits}

\begin{figure}[htb]
\begin{center}
\hbox{\hspace*{-0.15in}
\mbox{\epsfig{file=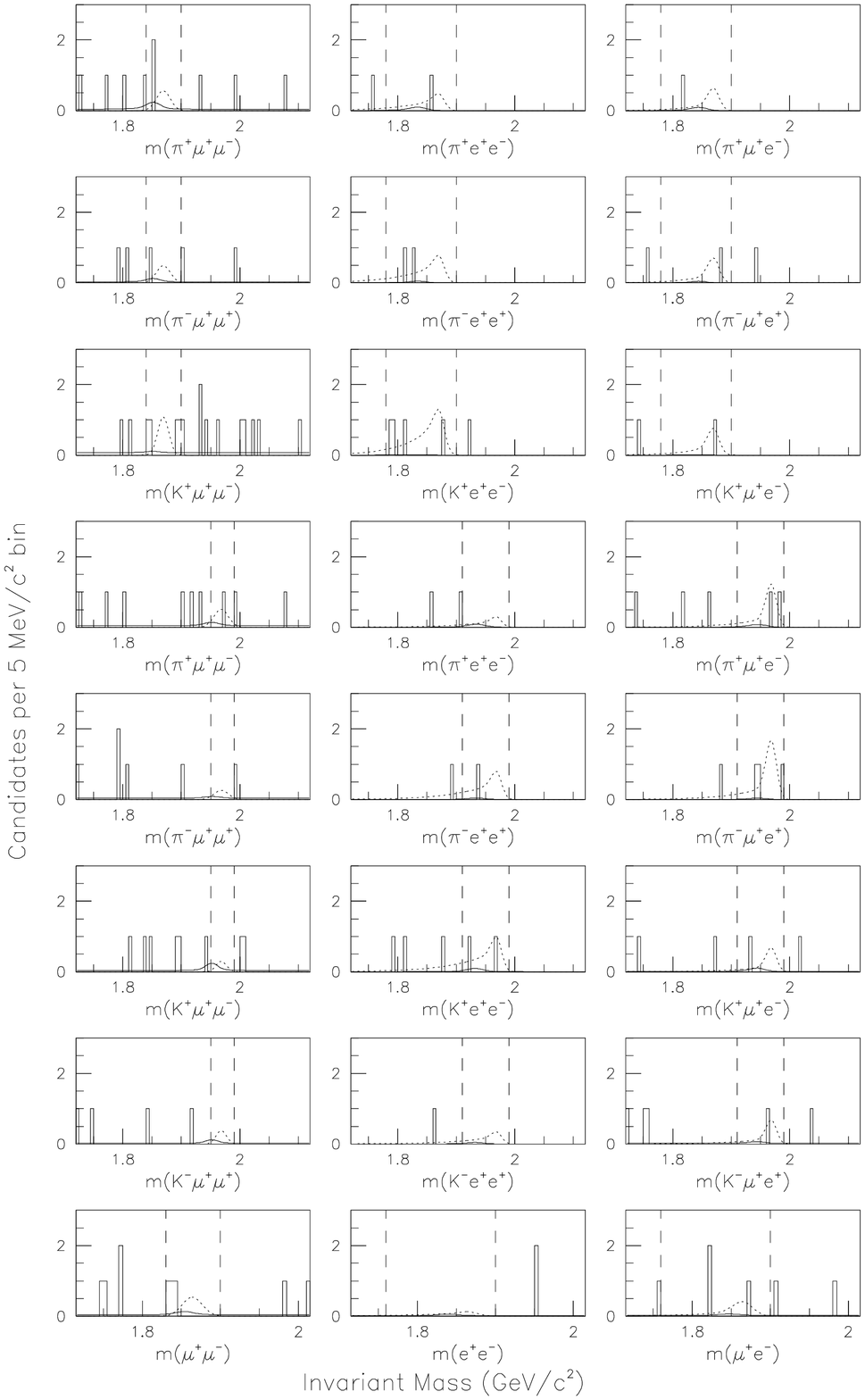,height=7.5in}}
}
\end{center}
\vspace*{-0.40in}
\caption{\it Final \dpll\ (rows 1--3), \dspll\ (rows 4--7),
and \dll\ (row 8) event samples. The solid curves represent 
estimated background; the dotted curves represent signal 
shape for an event yield equal to the 90\% C.L. upper limit; 
the dashed vertical lines denote the signal mass windows. 
\label{fig:openbox3body}}
\end{figure}

\begin{figure}[htb]
\begin{center}
\mbox{\epsfig{file=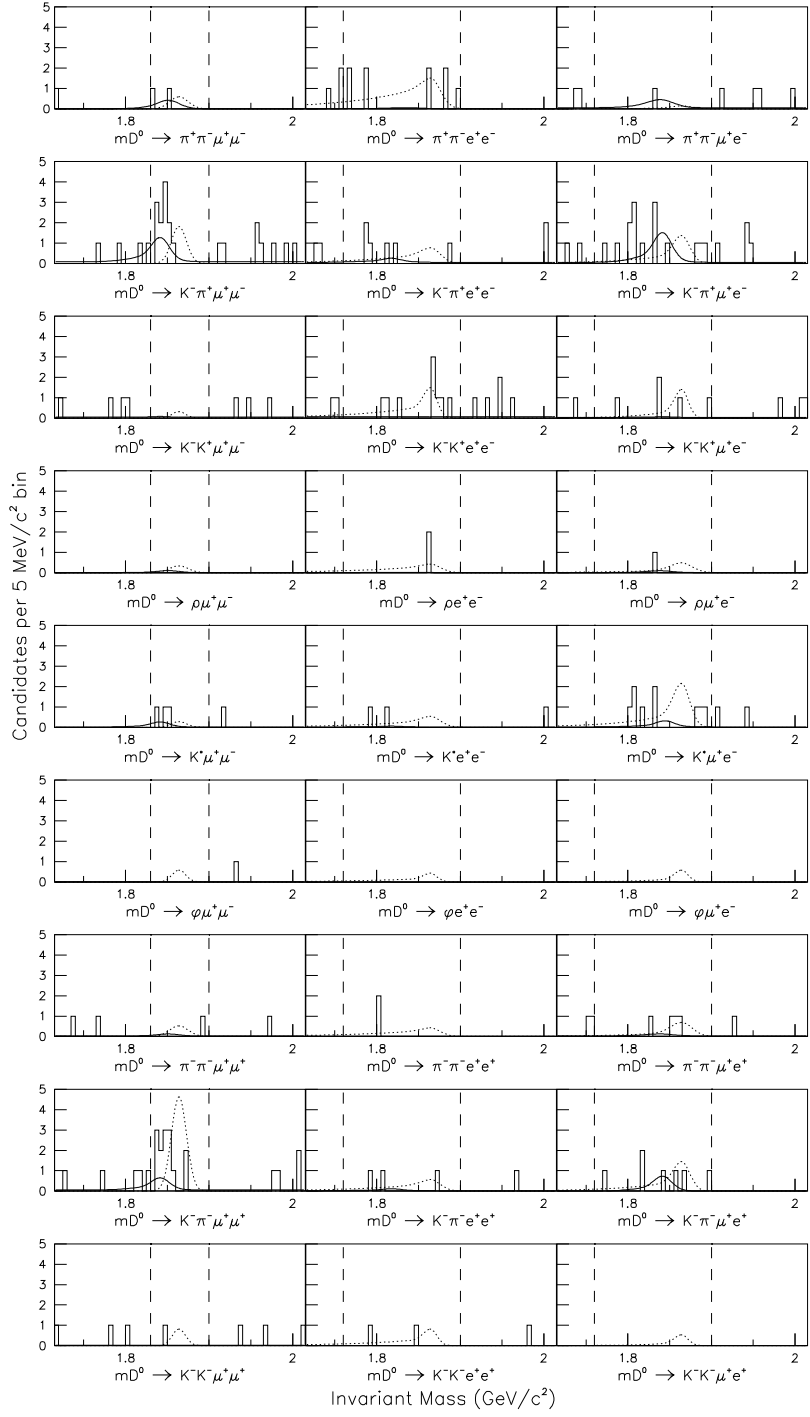,height=7.5in}}
\end{center}
\vspace*{-0.40in}
\caption{\it Final \dvll\ and \dhhll\ event samples 
for nonresonant modes (rows 1--3), resonant modes (rows 4--6), 
and same-signed dilepton modes (rows 7--9). The solid curves 
represent estimated background; the dotted curves represent 
signal shape for an event yield equal to the 90\% C.L. upper 
limit; the dashed vertical lines denote the signal mass windows. 
\label{fig:openbox4body}}
\end{figure}

\begin{figure}[htb]
\begin{center}
\mbox{\epsfig{file=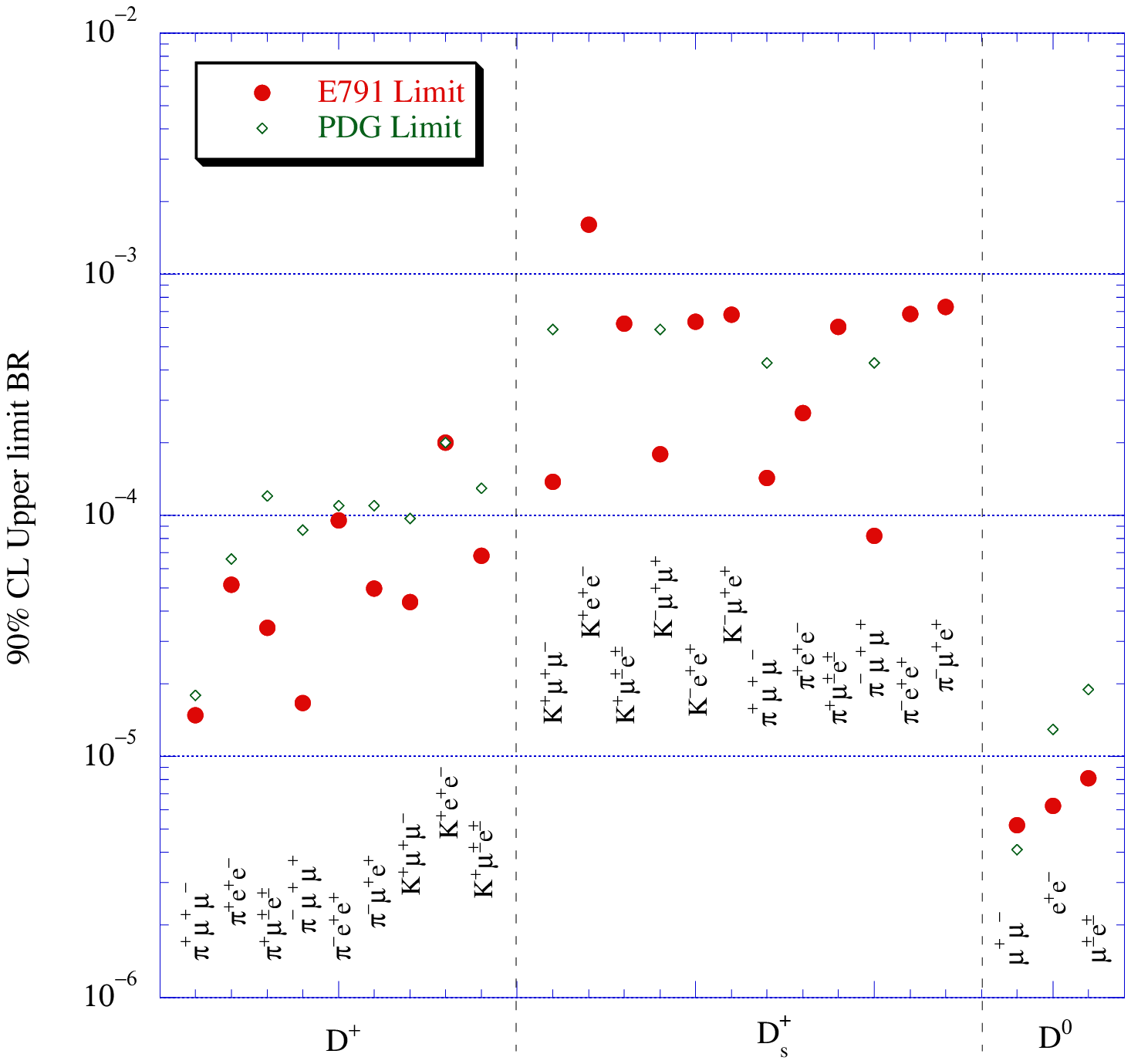,height=3.5in}}
\vskip0.20in
\mbox{\epsfig{file=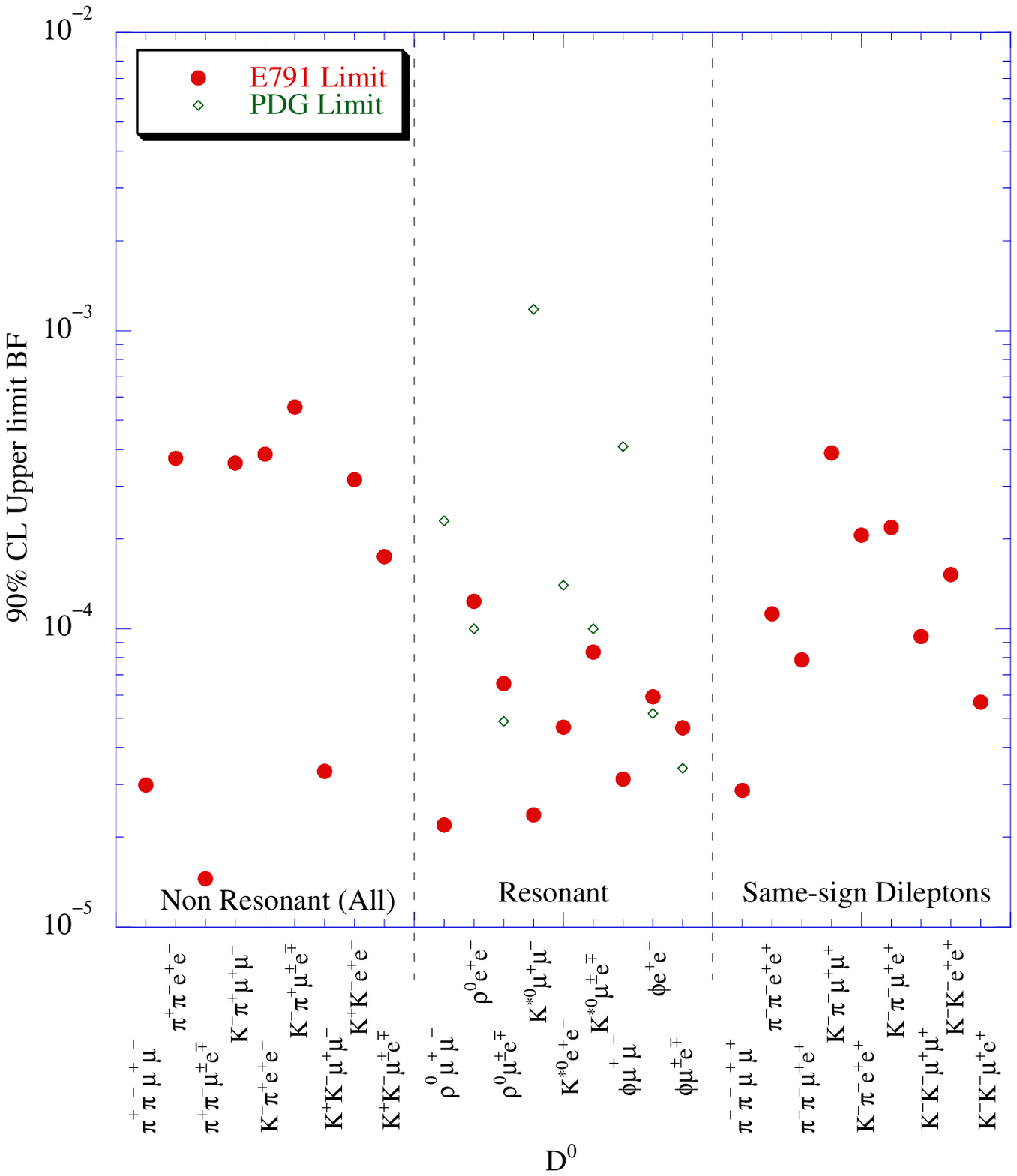,height=4.2in}}
\end{center}
\vspace*{-0.20in}
\caption{\it Upper limits for \dll\ and \dorspll\ decays (top),
and \dvll\ and \dhhll\ decays (bottom). The circles are E791 
limits\cite{e791:3body,e791:4body} and the diamonds are 
(previous) limits from the Particle Data Book\cite{PDG}.
\label{fig:results234body}}
\end{figure}

The final event samples after all selection criteria 
were applied and signal boxes ``opened'' are shown in 
Figs.~\ref{fig:openbox3body}--\ref{fig:openbox4body}.
The number of background events estimated, the number of 
candidate events observed, the overall systematic error, 
and the resultant 90\% C.L. upper limits are listed in 
Tables~\ref{tab:limits3body} and \ref{tab:limits4body}.
The systematic errors arise mainly from four sources: 
errors resulting from the fits to the normalization channels, 
statistical errors on the number of Monte Carlo events generated
and accepted (used to calculate acceptance), uncertainties in the 
amounts of reflection and combinatoric background, and uncertainties 
in the relative detection efficiencies between the search modes and 
their normalization channels. The upper limits on the branching
fractions are compared to (previous) limits from the Particle Data 
Book\cite{PDG} in Fig.~\ref{fig:results234body}. Of the 51 decay 
modes listed, all but six have upper limits more stringent than 
previously published results. For 26 of these modes, the E791 
limits are the first such limits reported.

\section{Other Experiments}

There are five published limits that remain superior to those 
obtained by E791, and one published limit that is equivalent.
These were obtained by the CLEO experiment\cite{cleo} 
($D^0\rightarrow\rho^0 e^+e^-,\ \rho^0\mu^\pm e^\mp,\ 
\phi\,e^+e^-,\ \phi\,\mu^\pm e^\mp$), by the
BEATRICE experiment\cite{beatrice} (\dmumu), and by
FNAL E687\cite{e687} (\dkee). This last experiment used a photon 
beam of mean energy $\sim$\,220\geve\ to photoproduce charm, 
and a silicon strip vertex detector (like E791) to reconstruct 
$D$ decay vertices. The experiment ran concurrently with E791 and 
obtained an upper limit for $B(D^+\rightarrow K^+e^+e^-)$ identical 
to that from E791. 

Another hadroproduction experiment, 
FNAL E771\cite{e771}, also ran concurrently with E791 but used an 
800\gevp\ proton beam. This experiment obtained an upper limit for 
$B(D^0\rightarrow\mu^+\mu^-)$ remarkably close to that obtained by 
BEATRICE. In fact, the 90\% C.L.\ upper limits from BEATRICE, E771, 
and E791 are within about 25\% of each other: 
$B(D^0\rightarrow\mu^+\mu^-)<4.1$, 4.2, and 5.2 $\times 10^{-6}$, 
respectively. Assuming these results uncorrelated, we combine them 
to obtain $B(D^0\rightarrow\mu^+\mu^-)<1.5\times 10^{-6}$ at 90\% C.L.
This limit is still many orders of magnitude larger than the Standard 
Model expectation\cite{Pakvasa} (dominated by long-distance effects) 
of~$\sim$\,$10^{-15}$.

For the future, we expect greater sensitivity than that of the above 
experiments from FNAL E831 (FOCUS)\cite{e831}, a photoproduction 
experiment that is an upgraded version of E687. The E831 detector 
employed more muon counters than did E791 and also used a finer-grained 
electromagnetic calorimeter. The experiment took data in 1996--97, 
recording a charm sample approximately four times larger than that 
of E791. The analysis of this data set is underway.

\section{Summary}

In summary, E791 has completed an extensive search for flavor-changing
neutral-current, lepton-flavor-violating, and lepton-number-violating
processes and sees no evidence for these decays. The experiment has 
set upper limits on 51 different decay modes; all limits but six 
are improvements over previously published results. Many of these 
limits are at the $10^{-5}$ to $10^{-6}$ level and can constrain 
various extensions\cite{LopezCastro} to the Standard Model. We 
anticipate even more stringent limits (or possibly signals) 
from E831, which has a ``cleaner'' data sample due to the 
photoproduction process, superior muon and electron 
identification, and approximately four times as many 
reconstructed charm decays.

\begin{table}[t]
\centering
\caption{ \it The number of background events estimated, the number 
of candidate events observed, the overall systematic error, and the 
90\% C.L. upper limits on $N^{}_X$ and the branching fraction 
for \dll\ and \dorspll\ decays.}
\vskip 0.1 in
\tabcolsep=4.5pt
\begin{tabular}{|c|ccc|cccc|}
\hline
 & & & & & & & \\
  &   \multicolumn{3}{c|}{\bf Est.\ background} &	& 
  {\bf System.} &    & {\bf BR} \\
  {\bf Mode} & $N^{}_{\rm comb}$ & $N^{}_{\rm misID}$ & $N^{}_{\rm tot}$ & 
  {\bf {\boldmath $N^{}_{\rm obs}$}} & {\bf {\boldmath Err.\,(\%)}} & 
  {\boldmath {\bf $N^{}_X$}} & {\bf {\boldmath $\times 10^{-5}$ }} \\
 & & & & & & & \\
\hline 
 & & & & & & & \\
$D^0\rightarrow\mu^+ \mu^-$ 
		& 1.83 & 0.63 &2.46 & 2  &  6 & 3.51 & 0.52 \\  
$D^0\rightarrow e^+ e^-$ 
		& 1.75 & 0.29 &2.04 & 0  &  9 & 1.26 & 0.62 \\  
$D^0\rightarrow\mu^\pm e^\mp$ 
		& 2.63 & 0.25 &2.88 & 2  &  7 & 3.09 & 0.81 \\  
 & & & & & & & \\
\hline
 & & & & & & & \\
$D^+\rightarrow\pi^+\mu^+ \mu^-$ 
		& 1.20 & 1.47 &2.67 & 2  & 10 & 3.35 & 1.5 \\  
$D^+\rightarrow\pi^+ e^+ e^-$ 
		& 0.00 & 0.90 &0.90 & 1  & 12 & 3.53 & 5.2 \\  
$D^+\rightarrow\pi^+\mu^\pm e^\mp$ 
		& 0.00 & 0.78 &0.78 & 1  & 11 & 3.64 & 3.4 \\  
$D^+\rightarrow\pi^-\mu^+ \mu^+$ 
		& 0.80 & 0.73 &1.53 & 1  & 9  & 2.92 & 1.7 \\  
$D^+\rightarrow\pi^- e^+ e^+$ 
		& 0.00 & 0.45 &0.45 & 2  & 12 & 5.60 & 9.6 \\  
$D^+\rightarrow\pi^-\mu^+ e^+$ 
		& 0.00 & 0.39 &0.39 & 1  & 11 & 4.05 & 5.0 \\  
$D^+\rightarrow K^+\mu^+ \mu^-$ 
		& 2.20 & 0.20 &2.40 & 3  &  8 & 5.07 & 4.4 \\  
$D^+\rightarrow K^+ e^+ e^-$ 
		& 0.00 & 0.09 &0.09 & 4  & 11 & 8.72 & 20 \\  
$D^+\rightarrow K^+\mu^\pm e^\mp$ 
		& 0.00 & 0.08 &0.08 & 1  &  9 & 4.34 & 6.8 \\  
 & & & & & & & \\
\hline
 & & & & & & & \\
$D^+_s\rightarrow K^+\mu^+ \mu^-$ 
		& 0.67 & 1.33 &2.00 & 0  & 27 & 1.32 & 14 \\  
$D^+_s\rightarrow K^+ e^+ e^-$ 
		& 0.00 & 0.85 &0.85 & 2  & 29 & 5.77 & 160 \\  
$D^+_s\rightarrow K^+\mu^\pm e^\mp$ 
		& 0.40 & 0.70 &1.10 & 1  & 27 & 3.57 & 63 \\  
$D^+_s\rightarrow K^-\mu^+ \mu^+$ 
		& 0.40 & 0.64 &1.04 & 0  & 26 & 1.68 & 18 \\  
$D^+_s\rightarrow K^- e^+ e^+$ 
		& 0.00 & 0.39 &0.39 & 0  & 28 & 2.22 & 63 \\  
$D^+_s\rightarrow K^-\mu^+ e^+$ 
		& 0.80 & 0.35 &1.15 & 1  & 27 & 3.53 & 68 \\  
$D^+_s\rightarrow\pi^+\mu^+ \mu^-$ 
		& 0.93 & 0.72 &1.65 & 1  & 27 & 3.02 & 14 \\  
$D^+_s\rightarrow\pi^+ e^+ e^-$ 
		& 0.00 & 0.83 &0.83 & 0  & 29 & 1.85 & 27 \\  
$D^+_s\rightarrow\pi^+\mu^\pm e^\mp$ 
		& 0.00 & 0.72 &0.72 & 2  & 30 & 6.01 & 61 \\  
$D^+_s\rightarrow \pi^-\mu^+ \mu^+$ 
		& 0.80 & 0.36 &1.16 & 0  & 27 & 1.60 & 8.2 \\  
$D^+_s\rightarrow \pi^- e^+ e^+$ 
		& 0.00 & 0.42 &0.42 & 1  & 29 & 4.44 & 69 \\  
$D^+_s\rightarrow \pi^-\mu^+ e^+$ 
		& 0.00 & 0.36 &0.36 & 3  & 28 & 8.21 & 73 \\  
 & & & & & & & \\
\hline
\end{tabular}
\label{tab:limits3body}
\end{table}

\begin{table}[t]
\centering
\caption{ \it The number of background events estimated, the number 
of candidate events observed, the overall systematic error, and the 
90\% C.L. upper limits on $N^{}_X$ and the branching fraction 
for \dvll\ and \dhhll\ decays. 
The total backround estimate ($N^{}_{tot}$) includes small
contributions from 
$D^0\rightarrow K^-\pi^+\rho^0$, 
$D^0\rightarrow K^+ K^-\rho^0$, 
$D^0\rightarrow\overline{K}^{*0}\rho^0$, and 
$D^0\rightarrow\phi\/\rho^0$, where
$\rho^0\rightarrow\ell^+\ell^-$.}
\vskip 0.1 in
\tabcolsep=5.5pt
\begin{tabular}{|c|ccc|cccc|}
\hline
 & & & & & & & \\
  &   \multicolumn{3}{c|}{\bf Est.\ background} &	& 
  {\bf System.} &    & {\bf BR} \\
  {\bf Mode} & $N^{}_{\rm comb}$ & $N^{}_{\rm misID}$ & $N^{}_{\rm tot}$ & 
  {\bf {\boldmath $N^{}_{\rm obs}$}} & {\bf {\boldmath Err.\,(\%)}} & 
  {\boldmath {\bf $N^{}_X$}} & {\bf {\boldmath $\times 10^{-5}$ }} \\
 & & & & & & & \\
\hline 
 & & & & & & & \\
$\pi^+\pi^-\mu^+ \mu^-$   & 0.00 & 3.16 &3.16 & 2  & 11 & 2.96 & 3.0 \\  
$\pi^+\pi^- e^+ e^-$      & 0.00 & 0.73 &0.73& 9  & 12 & 15.2 & 37 \\  
$\pi^+\pi^-\mu^\pm e^\mp$ & 5.25 & 3.46 &8.71& 1  & 15 & 1.06 & 1.5 \\  
$K^-\pi^+\mu^+ \mu^-$     & 3.65 & 0 & 4.10 & 12 & 11 & 15.4 & 36 \\  
$K^-\pi^+ e^+ e^-$        & 3.50 & 0 & 3.94 & 6  & 15 & 7.53 & 39 \\  
$K^-\pi^+\mu^\pm e^\mp$   & 5.25 & 0 & 5.25 & 15 & 12 & 17.3 & 55 \\  
$K^+ K^-\mu^+ \mu^-$      & 2.13 & 0.17 & 2.37 & 0  & 17 & 1.22 & 3.3 \\  
$K^+ K^- e^+ e^-$         & 6.13 & 0.04 & 6.23 & 9  & 18 & 9.61 & 32 \\  
$K^+ K^-\mu^\pm e^\mp$    & 3.50 & 0.17 &3.67& 5  & 17 & 6.61 & 18 \\  
 & & & & & & & \\
\hline
 & & & & & & & \\
$\rho^0\mu^+ \mu^-$              & 0.00 & 0.75 &0.75& 0  & 10 & 1.80 & 2.2 \\  
$\rho^0 e^+ e^-$                 & 0.00 & 0.18 &0.18& 1  & 12 & 4.28 & 12 \\  
$\rho^0\mu^\pm e^\mp$            & 0.00 & 0.82 &0.82& 1  & 11 & 3.60 & 6.6 \\  
$\overline{K}^{*0}\mu^+ \mu^-$   & 0.30 & 1.87 &2.43& 3  & 24 & 5.40 & 2.4 \\  
$\overline{K}^{*0} e^+ e^-$      & 0.88 & 0.49 &1.62& 2  & 25 & 4.68 & 4.7 \\  
$\overline{K}^{*0}\mu^\pm e^\mp$ & 1.75 & 2.30 &4.05& 9  & 24 & 12.8 & 8.3 \\  
$\phi\,\mu^+ \mu^-$                & 0.30 & 0.04 &0.35& 0  & 33 & 2.33 & 3.1 \\  
$\phi\,e^+ e^-$                   & 0.00 & 0.01 &0.01& 0  & 33 & 2.75 & 5.9 \\  
$\phi\,\mu^\pm e^\mp$              & 0.00 & 0.05 &0.05& 0  & 33 & 2.71 & 4.7 \\  
 & & & & & & & \\
\hline
 & & & & & & & \\
$\pi^-\pi^-\mu^+ \mu^+$   & 0.91 & 0.79 &1.70& 1  & 9  & 2.78 & 2.9 \\  
$\pi^-\pi^- e^+ e^+$      & 0.00 & 0.18 &0.18& 1  & 11 & 4.26 & 11 \\  
$\pi^-\pi^-\mu^+ e^+$     & 2.63 & 0.86 &3.49& 4  & 10 & 5.18 & 7.9 \\  
$K^-\pi^-\mu^+ \mu^+$     & 2.74 & 3.96 &6.69& 14 & 9  & 15.7 & 39 \\  
$K^-\pi^- e^+ e^+$        & 0.88 & 1.04 &1.91& 2  & 16 & 4.14 & 21 \\  
$K^-\pi^-\mu^+ e^+$       & 0.00 & 4.88 &4.88& 7  & 11 & 7.81 & 22 \\  
$K^- K^-\mu^+ \mu^+$      & 1.22 & 0.00 &1.22& 1  & 17 & 3.27 & 9.4 \\  
$K^- K^- e^+ e^+$         & 0.88 & 0.00 &0.88& 2  & 17 & 5.28 & 15 \\  
$K^- K^-\mu^+ e^+$        & 0.00 & 0.00 &0.00& 0  & 17 & 2.52 & 5.7 \\  
 & & & & & & & \\
\hline
\end{tabular}
\label{tab:limits4body}
\end{table}

\begin{thebibliography}{99}

\bibitem{e791:3body}	E. M. Aitala {\it et al.}, 
				Phys. Rev. Lett. {\bf 76}, 364 (1996);
			Phys. Lett. {\bf B462}, 401 (1999). 

\bibitem{e791:4body}	E. M. Aitala {\it et al.}, FERMILAB-Pub-00/280-E (2000),
			hep-ex/0011077 (submitted to Phys. Rev. Lett.). 

\bibitem{e791:epj}	E. M. Aitala {\it et al.,} 
			Eur. Phys. J.\,direct {\bf C4}, 1 (1999).

			See also: 
			E. M. Aitala {\it et al.,} 
			Phys. Lett. {\bf B371}, 157 (1996); 
			Phys. Rev. {\bf D57}, 13 (1998). 

\bibitem{PDG}		D. E. Groom {\it et al.\/} (Particle Data Group),
				Eur. Phys. J. {\bf C15}, 1 (2000); 
			C.~Caso {\it et al.\/} (Particle Data Group),
				Eur. Phys. J. {\bf C3}, 1 (1998).

\bibitem{FeldmanCousins} G. J. Feldman and R. D. Cousins, 
			Phys. Rev. {\bf D57}, 3873 (1998). 
 
\bibitem{CousinsHighland} R. D. Cousins and V. L. Highland,
			Nucl. Instrum. Meth. {\bf A320}, 331 (1992). 

\bibitem{cleo}		A. Freyberger {\it et al.,} 
			Phys. Rev. Lett. {\bf 76}, 3065 (1996);
			Phys. Rev. Lett. {\bf 77}, 2147 (1996).

\bibitem{beatrice}	M. Adamovich {\it et al.,} 
			Phys. Lett. {\bf B408}, 469 (1997). 

\bibitem{e687}		P. L. Frabetti {\it et al.,} 
			Phys. Lett. {\bf B398}, 239 (1997). 

\bibitem{e771}		T. Alexopoulos {\it et al.,} 
			Phys. Rev. Lett. {\bf 77}, 2380 (1996).

\bibitem{Pakvasa}	S. Pakvasa, UH-511-871-97 (1997), hep-ph/9705397.

\bibitem{e831}		P.~D.~Sheldon, Charm Mixing and Rare Decays, 
			in: Heavy Flavors~8, Proc.\ of the 8th
			Int.\ Symp.\ on Heavy Flavour Physics,
			Southampton, UK, 25-29 July 1999 
			(The Journal of High Energy Physics Conf. 
			Proceedings, eds. P.~Dauncey and C.~Sachrajda,
			{\tt http://jhep.sissa.it}).			

			See also:
			J. M. Link {\it et al.,} 
			Phys. Lett. {\bf B485}, 62 (2000); 
			Phys. Lett. {\bf B491}, 232 (2000).

\bibitem{LopezCastro}	See for example: G. L\'{o}pez Castro, 
			R. Mart\'{i}nez, and J. H. Mu\~{n}oz,
			Phys. Rev. {\bf D58}, 033003 (1998). 



\end{thebibliography}
\end{document}